\documentstyle[preprint,aps,floats]{revtex}
\tightenlines
\input psfig.tex
\begin{document}
\def\ba{\begin{eqnarray}}
\def\ea{\end{eqnarray}}
\def\be{\begin{equation}}
\def\ee{\end{equation}}
\def\({\left(}
\def\){\right)}
\def\[{\left[}
\def\]{\right]}
\def\lagrange {{\cal L}}
\def\del {\nabla}
\def\d {\partial}
\def\Tr{{\rm Tr}}
\def\half{{1\over 2}}
\def\fourth{{1\over 8}}
\def\bibi{\bibitem}
\def\S{{\cal S}}
\def\xx{\mbox{\boldmath $x$}}
\newcommand{\labeq}[1] {\label{eq:#1}}
\newcommand{\eqn}[1] {(\ref{eq:#1})}
\newcommand{\labfig}[1] {\label{fig:#1}}
\newcommand{\fig}[1] {\ref{fig:#1}}
\def\gsim{ \lower .75ex \hbox{$\sim$} \llap{\raise .27ex \hbox{$>$}} }
\def\lsim{ \lower .75ex \hbox{$\sim$} \llap{\raise .27ex \hbox{$<$}} }
\newcommand\bigdot[1] {\stackrel{\mbox{{\huge .}}}{#1}}
\newcommand\bigddot[1] {\stackrel{\mbox{{\huge ..}}}{#1}}
\title{Open Inflation, the Four Form and the Cosmological Constant}
\author{
Neil
Turok\thanks{email:N.G.Turok@damtp.cam.ac.uk}
and S.W. Hawking\thanks{email:S.W.Hawking@damtp.cam.ac.uk}}
\address{
DAMTP, Silver St, Cambridge, CB3 9EW, U.K.}
\date\today 
\maketitle
\begin{abstract}
Fundamental theories of quantum gravity such as supergravity
include a four form field strength which contributes to the cosmological
constant. The inclusion of such a field into our theory of
open inflation [1] 
allows an anthropic solution to the 
cosmological constant problem in which
cosmological
constant gives a small but non-negligible contribution to
the density of today's universe. We include a discussion of
the role of the singularity in our solution and a reply to Vilenkin's
recent criticism.
\end{abstract}

\section{Introduction}

Inflationary theory has for some time had two skeletons in its cupboard.
The first has been the question of the 
pre-inflationary initial conditions. The problem is to explain why
the scalar field driving inflation was initially
displaced from the true minimum of its effective potential.
One possibility is that this happened through a
supercooled phase transition, with the  field
being shifted away from its true minimum 
by thermal couplings. Another possibility is that the field became
trapped in a `false vacuum', a metastable
minimum of the potential. But 
both of these scenarios are hard to reconcile 
with the very flat potential and weak self-couplings
required to suppress the inflationary quantum fluctuations to
an acceptable level. 
Most commonly, people have simply placed
the field driving inflation high up its potential by hand in 
order to get inflation going. 
The problem here is that these initial conditions may be very unlikely.
The only proposed  measure on the space of initial conditions with some
pretensions to completeness, the Hartle-Hawking prescription for 
the Euclidean 
path integral \cite{HH}, predicts that inflationary initial conditions 
are exponentially improbable. 

The second problem for inflation is the cosmological constant. 
The effective cosmological constant is what drives inflation, so it
must be large during inflation. But it  
must also be cancelled to extreme accuracy after inflation to
allow the usual radiation and matter dominated eras.
With no explanation of how this cancellation could
occur, 
the practice has been to simply set the minimum of the
effective potential to be zero, or very nearly zero.
This is a terrible
fine tuning problem leading one to suspect that some important 
physics is missing.

In this paper we propose a solution to the cosmological constant problem, 
extending 
our recent paper on open inflation, where we
calculated the Euclidean path integral with the Hartle-Hawking 
prescription using a new family of
singular but finite action instanton solutions.
We found that in this approach
the simplest inflationary models
with a single scalar field coupled to gravity gave the 
unfortunate prediction that the most likely
open universes were nearly empty. 
We were forced to invoke the anthropic principle to determine the
value of $\Omega_0$. 
Imposing the minimal requirement that 
our galaxy formed led to the most probable
value for 
$\Omega_0$ being 0.01. This is far too 
low to fit current observations, although the issue is not 
completely straightforward because the region of gravitationally
condensed matter our galaxy would 
be in would necessarily be large, and would contain many other galaxies
\cite{HT}.

In this paper we extend the simplest scalar field models by including a
four form field, a natural addition to the Lagrangian which occurs
automatically in supergravity. 
The four form field's
peculiar properties 
have been known for some time: it provides a contribution to
the cosmological constant whose magnitude is not determined by the field
equations. This property was exploited before by one of us 
in an attempt to explain why the present cosmological constant might be 
zero
\cite{Hawkcos}. A subtlety in the calculation with the four form
was later pointed out by 
by Duff \cite{duff}, who showed that the Euclidean 
path integral
actually gave $\Lambda=0$ as the most {\it unlikely}
possibility. Here we shall perform the calculation appropriate to
an anthropic constraint on $\Lambda$ at late times. We 
shall show that in this context the four form allows an 
anthropic solution of the cosmological constant problem in which
the prior probability for $\Lambda$ is very nearly flat,
and the actual value
of $\Lambda$ today is then determined by considerations
of galaxy formation alone. 

An earlier version of this paper incorrectly claimed that 
Duff's calculation solved the empty universe problem. 
Bousso and Linde (private communication, \cite{linde}) pointed
out that the action we computed for the four form field was not
proportional to the geometric entropy. This prompted us to 
reconsider the calculation, and when we did so we discovered an error.
The problem with the calculation was that we used the action 
appropriate for computing the wavefunction in
the coordinate representation, whereas the anthropic constraint
on $\Lambda$ is a constraint on the momentum of the three form 
gauge potential. One therefore needs to compute the path integral 
for the wavefunction in 
the momentum representation, and this turns out to restore
the validity of Hawking's original result for the prior probability for
$\Lambda$. The empty universe problem 
remains, though there may be other solutions as were
mentioned in \cite{HT}, and will be discussed below.

In \cite{HT} we introduced a new family of singular but finite action
instantons which describe the beginning of inflationary universes.
Prior to our work the only known finite action
instantons were those which occurred when the scalar field
potential had a positive extremum \cite{HM} or a sharp false vacuum
\cite{CD}. In contrast, 
the family of instantons we found exists for essentially any 
scalar field potential. When analytically continued to the Lorentzian
region, the instantons describe infinite, open inflationary universes. 
Several subsequent papers have appeared, making various criticisms
of these instantons, and of our interpretation of them.
Linde \cite{linde} has made general arguments 
against the Hartle-Hawking prescription, to which we have replied in
\cite{HTcomm}. Vilenkin \cite{Vil} argues that singular instantons
must be forbidden or else they would lead to an instability of
Minkowski space. We respond to this criticism in Section III below.
Unruh \cite{unruh} has explored some of the properties
of our solutions and interpreted them in terms of a closed universe
including an ever growing region of an infinite
open universe. Finally,
Wu \cite{chao} has discussed interpreting 
instantons we use as `constrained' instantons.

The family of instantons we study
allows one
to compute the  theoretical prior probabilities for cosmological parameters 
such as the density parameter 
$\Omega_0$ and the cosmological
constant $\Lambda$ (where that is a free parameter, as it will be here)
directly from the
path integral for quantum gravity.
An interesting  consequence of our calculations 
is that over the range of values for the cosmological constant 
allowed by the anthropic principle, the theoretical prior probability
for $\Omega_\Lambda$ is very nearly flat. Thus there is a 
high probability that $\Omega_\Lambda$ is non-negligible in 
todays universe.

\section{The Four Form and The Euclidean Action}

The Euclidean action for the theory we consider is:
\be
{\cal S}_E = \int d^4 x \sqrt{g} \left( -{1\over 16 \pi G} R +{1\over 2}
g^{\mu \nu}\partial_\mu \phi \partial_\nu \phi + V(\phi) -{1\over 48}
F_{\mu \nu \rho \lambda} F^{\mu \nu \rho \lambda}\right) +\sum_i {\cal B}_i
\labeq{action}
\ee
where the sum includes surface terms which do not
contribute to the equations of motion, but are needed for the
reasons  to be explained. 
We use conventions where the Ricci scalar $R$ is positive for positively
curved manifolds. 
The inflaton field is 
$\phi$ and $V(\phi)$ is its scalar potential. The negative
sign of the $F^2$ term in the Euclidean action looks strange, but 
is actually implied by eleven dimensional supergravity compactified on 
a  seven sphere as described by Freund and Rubin \cite{FR}.
The minus sign is needed to reproduce the correct four dimensional 
field equations. The point is that the seven dimensional 
Ricci scalar contributes to the four dimensional Einstein equations,
with the contribution being proportional to the square of the
four form field strength  $F^2$, which determines the size
of the seven sphere.

The first surface term (which was neglected in \cite{HT}) 
occurs because
we wish to compute the path integral for the wavefunction of the 
three-metric in the 
coordinate representation. The 
Ricci scalar contains terms
involving  second derivatives of the metric, which 
are  undesirable because
when the action is
varied and one integrates by parts, they lead to
surface terms
involving normal derivatives of the metric variation on the boundary.
But the action we want is that relevant for computing the wavefunction in
the coordinate representation, and that should be stationary for arbitrary
variations of the metric which vanish on the boundary.

The second derivative terms can be
eliminated by integrating by parts, and the boundary term
turns out to be 
\be
{\cal B}_1 = \int d^3 x \sqrt{h} K/(8 \pi G)
\labeq{ba}
\ee
where
$K= h^{ij} K_{ij}$
is the trace of
the second fundamental form, calculated using the 
induced metric
$h_{ij}$ on the boundary
\cite{GH}.

The four form field strength
$F_{\mu \nu \rho \lambda}$ is
expressed in terms of its three-form potential as
\be
F_{\mu \nu \rho \lambda}= \partial_{[\mu}A_{\nu \rho \lambda]}.
\labeq{three}
\ee
The 
field equations for $F$, obtained by setting $\delta S /\delta 
A_{\nu\rho\lambda} =0$,
are 
\be
D_\mu F^{\mu \nu \rho \lambda}= {1\over \sqrt{g}} 
\partial_\mu(\sqrt{g} F^{\mu \nu \rho \lambda})
=0.
\labeq{threec}
\ee
The general solution is 
\be
F^{\mu \nu \rho \lambda}= {c\over i \sqrt{g}} \epsilon^{\mu \nu \rho \lambda}.
\labeq{threeba}
\ee
with $c$ an arbitrary constant, and where we have inserted a 
factor of $i$ so that the four form will be real in the Lorentzian 
region. 

The quantity $\sqrt{g} F^{0123}$ is 
the canonical momentum conjugate to the three form potential
$A_{123}$. The four form theory has no propagating degrees of freedom:
its only degree of freedom is the constant $c$ which corresponds to 
the momentum $p$ of a free particle  in one dimension.
As we shall see below, the constant $c$ is what determines
the cosmological constant today, and we shall be imposing
an anthropic constraint on that. So we want to compute the 
wavefunction as a function of the canonical momentum $\sqrt{g}F^{0123}$, not 
the coordinate $A_{123}$. (There was an error in the earlier
version of this paper on this point - for analogous considerations
regarding black hole duality see \cite{HR}). The action 
relevant for computing the wavefunction in the momentum 
representation should be stationary under arbitrary variations
which leave the momentum $F_{0123}$ unchanged on the boundary.
This action is obtained by adding a boundary term which cancels the
dependence on the variation of the gauge field $\delta A_{\nu \rho \lambda}$
on the boundary. 
The variation of the modified action then 
equals a term involving the 
the equations of motion plus a term proportional to
$\delta F_{0123}$ evaluated on the boundary, which is zero. 
The required boundary term is
\be
{\cal B}_2 = - \int d^3 x \sqrt{h} {1\over 24}
F^{\mu \nu \rho \lambda} A_{\nu \rho \lambda} n_\mu
\labeq{bb}
\ee
where $n^\mu$ is the unit vector normal to the boundary.
This term may be rewritten as the integral of a  total divergence:
\be
{\cal B}_2 = -\int d^4 x {1\over 24} \partial_\mu\left(
\sqrt{g} 
F^{\mu \nu \rho \lambda} A_{\nu \rho \lambda} \right).
\labeq{bbdiv}
\ee
When this term is evaluated on a solution to the field
equations (\ref{eq:threec}), it equals precisely minus twice
the original $\int \sqrt{g} {1\over 48} F^2$ term.

In the Lorentzian region (where  $g$ is negative) this solution continues to
\be
F^{\mu \nu \rho \lambda}= {c\over \sqrt{-g}} \epsilon^{\mu \nu \rho \lambda}
\labeq{threea}
\ee
which is real for real $c$. Note that the quantity 
\be
F^2 = F^{\mu \nu \rho \lambda} F_{\mu \nu \rho \lambda} =
-24 c^2 
\labeq{threeb}
\ee
is constant and real 
in both the Euclidean and Lorentzian regions.

The Einstein equations,  given 
by setting  $\delta S /\delta g_{\mu \nu} =0$, are
\be
G_{\mu \nu}= 8 \pi G \left[T_{\mu \nu}^\phi -{1\over 6}
 \left(F_{\mu \alpha \beta \gamma} F_{\nu}^{\, \alpha \beta \gamma} -{1\over 8} g_{\mu \nu} F^{\alpha \beta
 \gamma \delta}  F_{\alpha \beta
 \gamma \delta}\right)\right],
\labeq{eins}
\ee
with $T_{\mu \nu}^\phi$ the stress energy of the scalar field.
Taking the trace of this equation one finds
\be
R= 8\pi G\left( (\partial \phi)^2 +4V(\phi) + {1\over 12} F^2 \right),
\labeq{trace}
\ee
so that from (\ref{eq:action}), (\ref{eq:ba}) and (\ref{eq:bbdiv}) 
the Euclidean action is just
\be
{\cal S}_E =- \int d^4 x \sqrt{g} \left(V(\phi) +{1\over 48} F^2 \right)
+{1\over 8 \pi G}
\int d^3 x \sqrt{h} K . 
\labeq{actfin}
\ee

Now we follow our previous work in looking for $O(4)$ invariant solutions
to the Euclidean field equations. The four form field does not
contribute to the scalar field equations of motion, so the
solutions are just those we found before \cite{HT}, but with the
constant term ${1\over 48} F^2$ added to the scalar field potential
in the Einstein equations. 

The instanton metric is given in the Euclidean region by
\be
ds^2= d\sigma^2 +b^2(\sigma)d \Omega_3^2
\labeq{metric}
\ee
with $d \Omega_3^2$ the metric for the three sphere, and $b(\sigma)$ 
the radius of the three sphere. The field equation for the 
scalar field is
\be
\phi''+3 {b'\over b} \phi' =V_{,\phi},
\labeq{feq}
\ee
and the Einstein constraint equation is
\be
\left({b'\over b}\right)^2 = {1\over 3 M_{Pl}^2} \left({1\over 2} \phi'^2 -V_F
\right)
+{1\over b^2}
\labeq{fmet}
\ee
where $V_F=V+{1\over 48} F^2$ and primes denote derivatives with respect to
$\sigma$.
The instantons discussed in \cite{HT} are solutions to these
equations in which $b =\sigma +o(\sigma^3)$ and $\phi=\phi_0 +o(\sigma^2)$
near $\sigma=0$. As $\sigma$ increases there is a singularity, where $b$
vanishes as  $(\sigma_{f}-\sigma)^{1\over 3}$, and $\phi$ diverges
logarithmically. The Ricci scalar diverges at the singularity
as ${2\over 3} (\sigma_f-\sigma)^{-2}$. 

The presence of the singularity at the south pole of the 
deformed four sphere means that to evaluate the instanton 
action we  have to 
include the surface term evaluated 
on a small three sphere around the south pole.
The surface term in the action is calculated by noting that
the action density involves $\sqrt{g} R = -6(b''b+b'^2 -1)b$.
The second derivative
term can be integrated by parts to produce an action with first derivatives 
only. Doing so produces a surface term which must be cancelled by
the boundary term above. The required boundary term is thus
\be
{1\over 8 \pi G}
\int d^3 x \sqrt{h} K = - {1\over 8 \pi G} (b^3)' \int d \Omega^3 
\labeq{boundary}
\ee
where $\int d \Omega^3=\pi^2$ is half the volume of the three sphere.

The complete Euclidean instanton action is given by
\be
{\cal S}_E = - \pi^2 \int_0^{\sigma_f}  d\sigma b^3(\sigma) V_F(\phi) 
- \pi^2 M_{Pl}^2 (b^3)'(\sigma_f)
\labeq{fullact}
\ee
with $M_{Pl}= (8\pi G)^{-{1\over 2}}$ the reduced Planck mass.

For the flat potentials of interest, a good approximation to
the volume term is obtained by treating $V(\phi)$ as
constant over most of the instanton.
The surface term can be rewritten as a volume integral 
over $ V_{,\phi}$ as follows. Near the boundary of
the instanton, 
the gradient term $\phi'^2$ dominates over the potential
and the Einstein constraint equation (\ref{eq:fmet}) 
yields $b'\approx \phi'b/(\sqrt{6}M_{Pl})$.
We then rewrite the surface term (\ref{eq:boundary}) as
\be 
M_{Pl}^2(b^3)'(\sigma_f)=3 M_{Pl}^2 b^2b'(\sigma_f)\approx \sqrt{3\over 2} M_{Pl} 
b^3 \phi'(\sigma_f) = \sqrt{3\over 2} 
\int_0^{\sigma_f} d\sigma b^3(\sigma) M_{Pl} V_{,\phi}. 
\labeq{surfeval}
\ee
where we used the scalar field equation (\ref{eq:feq}) 
in the last step.
We perform the integral by treating $ V_{,\phi}$ as
constant. The integral is performed 
using the approximate solution 
$b(\sigma) \approx H^{-1} {\rm sin} (H \sigma)$, 
where $H^2= V_F/(3 M_{Pl}^2)$.
One finds $\int_0^\pi d \sigma b^3(\sigma)
\approx {4\over 3} H^{-4} = 12 M_{Pl}^4/V_F^2$.

With these approximations the
Euclidean action 
(\ref{eq:actfin}) is given by
\be
{\cal S}_E \approx - 12 \pi^2  M_{Pl}^4 
\left[ {1\over V_F(\phi_0)}
-{\sqrt{3\over 2} M_{Pl} V_{,\phi}(\phi_0)) \over 
V_F^2(\phi_0)} \right]
\labeq{actval}
\ee
where $\phi_0$ is the initial scalar field value, and the term 
containing $ V_{,\phi}(\phi_0)$ is the surface contribution.

Before continuing, we must deal with the issues of principle raised
by the existence of the singularity.  

\section{Avoiding the Singularity}

One might worry that the presence of a singularity meant that one could not use 
the instanton to make sensible physical predictions \cite{Vil} but this 
is not the case. The important point is that to calculate a wave function one 
only needs half an instanton \cite{chao}.  In other
words, the wave function $\Psi [h_{ij},\phi ]$ 
for a metric $h_{ij}$ and matter 
fields $\phi $ on a three surface $\Sigma $ is given by a path integral over 
metrics and matter fields on a four
 manifold $B$ whose only boundary is $\Sigma$.
We shall assume that the dominant contribution to this path integral comes 
from a non singular solution of the field equations on $B$. Then the 
probability of finding $ h_{ij} $ and $\phi $ on $\Sigma $ is
\be 
| \Psi |^2
\labeq{psi}
\ee
This can be represented by the double of $B$, that is, two copies of $B$ joined 
along $\Sigma $. Only in exceptional cases will the double be smooth on $\Sigma 
$. In general
if one analytically continues the solution on one $B$ onto the 
other it will have singularities.

Because one is interested
 in the probabilities for Lorentzian spacetimes, one has 
to impose the Lorentzian condition \cite{Lorentzian}
\be
Re(\pi^{ij})=0
\labeq{pi}
\ee
where $\pi^{ij} $ is the Euclidean momentum conjugate to $ h_{ij} $. This 
condition ensures that the second
fundamental form of $\Sigma$ is imaginary, that is, Lorentzian. One way of 
satisfying this condition in the solution considered in \cite{HT}
is to continue the 
coordinate $\sigma$ as $\sigma =\sigma_e + it $ where $\sigma _e $ is the 
value at the equator where the radius $b(\sigma )$ of the three spheres is 
maximal. This gives the wave function for a closed homogeneous and isotropic 
universe. In this case $B$ can be taken to be the Euclidean region from the 
north pole to the equator plus this Lorentzian continuation in imaginary 
$\sigma $. Clearly this is non singular since it doesn't include the south 
pole.

There is another way of slicing our $O(4)$ solution with a three surface 
$\Sigma $ of zero second fundamental form: a great circle through the north and 
south poles. Let $\chi $ be a coordinate on the instanton which is zero on the 
great circle but with non zero derivative. Then $t=i\chi $ will be a Lorentzian 
time and the surfaces of constant $t$ will be inhomogeneous
three spheres that sweep out a deformed de Sitter like solution.
The light cone of the north pole of the $t=0$ surface will contain the open 
inflationary universe and there will be a time like singularity running through 
the south pole. One might think this singularity would destroy one's ability to 
predict because the Einstein equations  do not hold there. However one can 
deform $\Sigma $ in a small half three sphere on one side of the singularity at 
the south pole and take $B$ to be the region on the non singular side of 
$\Sigma $. The deformation of $\Sigma $ near the south pole means that the 
Lorentzian condition will not be satisfied there. However this does not matter 
because this is not in the open universe region where observations of the 
Lorentzian condition are made. This is the important difference with the 
asymptotically flat singular instantons considered by Vilenkin \cite{Vil}
 in which the 
singularity expands to infinity and 
would be in the region of observation. The 
double of $B $ will be the whole $O(4)$ solution apart from a small
region round the south pole. One therefore has to include a surface
term at the 
south pole, as we have done above.

\section{The value of $\Lambda$ and $\Omega_0$
}

Let us consider a scalar field potential 
\be
V(\phi)= V_0 +V_1(\phi); \qquad {\rm min} \quad V_1(\phi)\equiv 0.
\labeq{pot}
\ee
so that $V_0$ represents the minimum potential energy. We
shall assume that $V_1$ is monotonically increasing over the range of
initial fields $\phi_0$ of interest.
In most inflationary models $V_0$ is simply set 
to zero by hand. Here the $F$ field can be chosen to 
cancel the `bare' cosmological constant. This could occur
for some symmetry or dynamical reason which we do not yet
understand, or for anthropic reasons as we
discuss below.

For the moment let us just assume that the $F$ field is chosen such that
the effective cosmological constant today vanishes. This 
condition reads
\be
\Lambda= V_0+{1\over 48} F^2 =0.
\labeq{lamo}
\ee
If $V_0$ is positive 
this requires real $F$ in the Lorentzian region, 
and imaginary $F$ in the Euclidean region. From the 
point of view of eleven dimensional supergravity,
including a positive $V_0$ cancels the negative
four dimensional cosmological constant of the Freund-Rubin 
solution, allowing a four dimensional universe with zero cosmological
constant. (The Freund-Rubin solution gives four dimensional anti-De Sitter
space cross a seven sphere). The condition that $V_0$ be positive is 
very interesting in the light of the well known fact that 
this is a requirement for supersymmetry breaking. 
Another implication of (\ref{eq:lamo}) is that the radius
of the seven dimensional sphere is  $R \sim M_{Pl}/V_0^{1\over 2}$.

Substituting  (\ref{eq:lamo}) back into the Euclidean action, 
we find 
\be
{\cal S}_E \approx - 12 \pi^2  M_{Pl}^4 \left( {1\over V_1(\phi_0)}
- { \sqrt{3\over 2} 
M_{Pl}V_{1,\phi}(\phi_0))\over V_1(\phi_0)^2} 
\right)
\labeq{actvalfsub}
\ee
where we now have 
terms of opposite sign contributing to ${\cal S}_E$. For example
if $V_1(\phi)\propto \phi^2$, the first term
goes $- \phi_0^{-2}$ whereas the second
goes as $+\phi_0^{-3}$. So the minimum Euclidean 
action occurs at some nonzero value of $\phi_0$, just what we
need for inflation\cite{BL}. However for general polynomial 
potentials it is straightforward to check that this effect
is not enough to give much inflation \cite{BL}.

However, for a potential with a local maximum, such as
 $V_1=\mu^4(1-{\rm cos}(\phi/v))$, 
one obtains a second local minimum of the Euclidean action at the 
maximum of the potential. The point is that 
if we expand about the maximum, in this case $\phi_0= v(\pi-\delta)$ with 
$\delta$ small, 
then the $V_{1,\phi}$ contribution 
to the Euclidean action 
increases linearly with $\delta$, whereas $V_1$ itself includes
only quadratic corrections in $\delta$. Therefore $\delta=0$ is a local
minimum of the Euclidean action. Consider the case
$v/M_{pl} >> 1$, $\mu <<M_{Pl}$,  so that the potential is very flat. 
As $\delta$ increases away from zero, $V_1$ decreases and the action
turns over, becoming smaller than the value at  $\delta=0$ when
$\delta \sim \sqrt{6} M_{Pl}/v$. Universes with $\delta$ larger than this
have a larger prior probability. 
But the number of inflationary efoldings
$N \approx M_{Pl}^{-2} \int_0^{\phi_0} d\phi (V_1/V_{1,\phi})
\approx 2 (v/M_{Pl})^2 {\rm log}(1/\delta)$.
For example if
$ v^2/M_{Pl}^2 \sim 10$, 
the number of efoldings corresponding to
$\delta > \sqrt{6} M_{Pl}/v$ would be small, and 
the corresponding universes would be much 
too open to allow galaxy formation. So one can concentrate on
the region around $\delta=0$. 
The problem with very small $\delta$ is that
the density perturbation amplitude 
$\Delta^2 = V_1^3/(M_{Pl}^6 V_{1,\phi}^2) \approx 
8 \mu^4 v^2/(M_{pl}^6 \delta^2)$ is
very large. Such universes might also be 
ruled out by anthropic considerations,
for a recent discussion see
\cite{TR}. The latter 
authors argue that 
if $\Delta^2$ is only modestly larger than
the value set by normalising to COBE, 
one would form galaxies so dense that
planetary systems would be impossible. 
This consideration disfavours $\delta$ being too small. 
Whether the anthropic effect is strong enough to counteract the Euclidean action
remains to be seen.

\section{The Anthropic Fix for $\Lambda$}

Now let us return to the cosmological constant. Since we do not at present
have any physics reason for the $F$ field to cancel the bare cosmological
constant, we resort to an anthropic argument. As Weinberg \cite{weinberg}
points out, anthropic arguments are particularly powerful when applied
to the cosmological constant, because there is a convincing case that 
unless the cosmological constant today is extremely small in Planck units, 
the formation of life would have been impossible. A very important
and perhaps even compelling
feature of the
anthropic argument is that it applies to the full cosmological 
constant, after all the contributions from electroweak symmetry breaking,
confinement and chiral symmetry breaking have been taken into account. 

The expression (\ref{eq:actval}) 
gives us the theoretical prior probability ${\cal P} (\phi_0,F^2)
\sim  e^{-2 {\cal S}_E(\phi_0,F^2)}$
for the four form
$F^2$ and the initial scalar field $\phi_0$. But
most of the possible universes 
have large positive or negative 
cosmological constants, and life would be impossible in them.
Following \cite{HT}, we shall assume what seems the
minimal conditions needed for our existence,
namely that our galaxy formed and lasted long enough for 
life to evolve. The latter condition eliminates large negative values of 
$\Lambda$, since the universe would have recollapsed too soon.
Large positive values for $\Lambda$ are excluded
because
$\Lambda$ domination would occur during  the 
radiation epoch, before the galaxy scale could re-enter the
Hubble radius. This would drive a second phase of 
inflation, which would never end. 
These two conditions alone force
$\Lambda$ to be very small in Planck units. 
Note that since the fluctuations are approximately scale invariant in
the theories of interest, the precise definition of a `galaxy'
is unimportant. The broad conclusions we reach here would apply even if we
took the `galaxy' mass scale to be as small as a solar mass.

We implement the anthropic principle via 
Bayes theorem, which  tells us that the posterior probability
for $\phi_0$ and $F^2$ is given by 
\be
{\cal P} (\phi_0,F^2|{\rm gal}) \propto {\cal P} ({\rm gal}|\phi_0,F^2)
 {\cal P} (\phi_0,F^2)
\labeq{prob}
\ee
where first factor represents the probability that a galaxy sized
region about us underwent gravitational collapse, given 
$\phi_0$ and $F^2$,  and the second
is the theoretical prior probability ${\cal P} (\phi_0,F^2)
\sim  e^{-2 {\cal S}_E(\phi_0,F^2)}$.
We want to maximise (\ref{eq:prob}) as a function of the 
initial field $\phi_0$ and the four form field $F^2$, 
or equivalently of $\Omega_0
=\Omega_M+\Omega_\Lambda$ and 
$\Omega_\Lambda$. 

Consider the $\Omega_\Lambda$ dependence of  (\ref{eq:actval}) first. 
The galaxy formation probability
${\cal P} ({\rm gal}|\phi_0,F^2)$ 
is negligible unless 
$\Lambda$ domination happened after the galaxy scale 
re-entered the Hubble radius, at $t \sim 10^9$ seconds.
We re-express $\Lambda$ as $\Lambda=\Omega_\Lambda \rho_c$ where
$\rho_c= 3 H_0^2/(8 \pi G) = 3 H_0^2 M_{Pl}^2$ is the critical
density. The condition that $\Lambda$ domination happened later than
$10^9$ seconds after the big bang 
reads
$|\Omega_\Lambda| < 10^{17}$, a mild constraint but 
strong enough for us to draw an important conclusion.
We expand the Euclidean action in $\Omega_{\Lambda}$ to obtain
\be
{\cal S}_E = 12 \pi^2 M_{Pl}^4 \bigl[-{1 \over V_1} \left(
1-6 {\Omega_\Lambda M_{Pl}^2 H_0^2 \over V_1}\right) -
{9 \Omega_\Lambda M_{Pl}^2 H_0^2 \over V_1^2} +..\bigr].
\labeq{exact}
\ee
The point is that the present Hubble constant
$H_0$ is {\it tiny} compared to $V_1$: in the example above 
we had 
$V_1(\phi_0) \approx 120 M_{Pl}^2 m^2$, and normalising to COBE requires
$m^2 \approx 10^{-11} M_{Pl}^2$. But today's Hubble constant is
$H_0 \sim 10^{-60} M_{Pl}$, so that 
even the above very minimal bound on $\Omega_\Lambda$ means
that the quantity we are expanding in, 
$H_0^2 M_{Pl}^2 \Omega_\Lambda / V_1 < 10^{-94}$!
Thus over the range of values of 
of $F^2$ such that we can even {\it discuss}
 the possibility of galaxies existing, 
the dependence of the Euclidean action on $\Omega_\Lambda$ is
completely negligible. 

Likewise, if a physical mechanism such as 
the cosine potential described above 
increases $\phi_0$ 
so that 
we get an acceptable value $0.1<\Omega_0<1.0$ today,
the $\phi_0$ dependence of the prior probability is likely to 
massively  outweigh that 
of the galaxy formation probability. The reason for this is
the Euclidean action depends inversely on $V_1(\phi_0)$. 
If we are to match COBE, $V_1(\phi_0)$ has to be much smaller than
the Planck density and the Euclidean action is enormous.
However, if we normalise to COBE and $\Omega_0$ is
not far from unity, the
galaxy formation probability is a function of $\Omega_0$ 
containing no 
large dimensionless number.  
So the problem of maximising the joint 
probability factorises. The anthropic principle 
fixes $\Lambda$ to be small, and the Euclidean action (or prior probability)
then fixes $\Omega_0$.

One can also consider 
the posterior probability for $\Lambda$ within this framework. As we have 
argued, the  posterior probability is to a good approximation completely determined 
by the galaxy formation probability alone. The possibility that
this might be the case was 
anticipated by 
Weinberg\cite{weinberg} and Efstathiou \cite{efstathiou}.

Let us briefly review the effect on galaxy formation of varying 
$\Lambda$, for modest values of $\Omega_\Lambda$ today.
In (\ref{eq:prob}), we should use
\be
{\cal P} ({\rm gal}|\phi_0,F^2) \sim {\rm erfc} (\delta_c /\sigma_{gal})
\labeq{gal}
\ee
where we assume Gaussian statistics. 
Here, $\delta_c$ is the value of the linear density perturbation 
required for gravitational collapse, usually taken to be that in the
spherical collapse model, 
$\delta_c = 1.68$. 
The amplitude of 
density perturbations on the galaxy scale in 
today's  universe, $\sigma_{gal}$ is 
given roughly by
\be
\sigma_{gal}\approx \Delta(\phi_{gal}) G(\Omega_M,\Omega_\Lambda) 
\labeq{sig}
\ee
where $\Delta(\phi_{gal})\sim 3 \times 10^{-4}$ 
is the amplitude of perturbations 
at horizon crossing, fixed by normalising to  COBE, 
and $G(\Omega_M,\Omega_\Lambda)$
is the growth factor for density perturbations in the
matter era. The latter varies strongly with $\Omega_M$: for example
in a flat universe, with $\Omega_\Lambda=1-\Omega_M$, we have 
$G \propto \Omega_M^{10\over 7} = (1-\Omega_\Lambda)^{10\over 7}$ 
at small $\Omega_M$ \cite{lahav},
whereas in an open universe with small $\Omega_\Lambda$ 
we have  $G\propto \Omega_M^2 \propto (\Omega_0 -\Omega_\Lambda)^2$.
One factor of $\Omega_M$ occurs because of the
change in the redshift of matter-radiation equality, and the 
remaining dependence is 
due to  the loss of growth at 
late times. In any case, for fixed $\phi_0$ and therefore fixed 
total density $\Omega_M+\Omega_\Lambda$, reducing $\Omega_\Lambda$
increases the probability of galaxy formation. So 
for fixed $T_0$ and $H_0$
the most probable
value of $\Lambda$ is zero, but
there is a high probability for non-negligible $\Omega_\Lambda$. 
Detailed computations of the posterior probability for $\Omega_\Lambda$
have been carried out by Efstathiou \cite{efstathiou} and 
Martel et al. \cite{martel}. It would be interesting to generalise
these to the open universes discussed here.

\section{Conclusions}

We have reached the somewhat surprising conclusion that
the universe most favoured by simple inflationary models 
with a four form field is open and with a small but non-negligible 
cosmological constant today. Our use of the anthropic argument to
fix $\Lambda$ is not new, and the possibility that the theoretical prior
probability might be a very flat function of $\Omega_\Lambda$ was
anticipated.
However it is an important advance that we can 
actually calculate the prior probability from first 
principles. 

Finally, we emphasise that the problem of explaining 
why $\Omega_0>0.01$ today remains, although we have 
noticed some promising aspects of potentials with
local maxima in this regard. As we have mentioned, in 
that case the problem is to understand whether 
anthropic considerations disfavour very large perturbation amplitudes
as strongly as the Euclidean action favours the initial field
starting near the potential maximum.

\medskip
\centerline{\bf Acknowledgements}
We thank R. Bousso, R. Crittenden, G. Efstathiou,
S. Gratton, A Linde and H. Reall for useful discussions
and correspondence. We especially acknowledge 
Bousso and Linde for pointing out the discrepancy between 
our original calculation of the action and the
geometric entropy.

\end{document}